\documentclass[review,jog]{igs}

\makeatletter
\def\@journal{}
\makeatother

\usepackage[utf8]{inputenc}
\usepackage{mathabx}
\usepackage{graphicx}
\usepackage{siunitx}
\usepackage{subfig}
\usepackage{url}
\usepackage{multirow}

\usepackage{hyperref}
\hypersetup{hidelinks}

\usepackage{xcolor}
\AddToHookNext{shipout/foreground}{%
  \put(0.5\paperwidth,-1.8cm){\makebox[0pt]{\footnotesize\sffamily\color{red!40!gray}%
    PREPRINT: this manuscript has been submitted to \textit{Journal of Glaciology} and has not yet been peer reviewed.}}}

\titleformat{\section}{\large\bfseries\raggedright}{\thesection}{1em}{\MakeUppercase{#1}}
\titleformat{\subsection}{\large\bfseries\raggedright}{\thesubsection}{1em}{#1}
\titleformat{\subsubsection}{\itshape\raggedright}{\thesubsubsection}{1em}{#1}

\jourvolume{xx}
\jourissue{x}
\jourpubyear{xxx}

\begin{document}

\nolinenumbers
\linespread{1}\selectfont

\title[Sub-Ice]{Sub-Ice: semi-automated mapping of basal channels from ice shelf surface elevation data}

\author[Kreynen and others]{Dylan Kreynen$^1$, Karen E. Alley$^2$, Marcelo Santis$^{3,4}$, Regine Hock$^{1,5}$, Phoebe H. Jackson$^6$}

\affiliation{%
$^1$Department of Geosciences, University of Oslo, Oslo, Norway\\
$^2$Centre for Earth Observation Science, University of Manitoba, Winnipeg, MB, Canada\\
$^3$Department of Geology, University of Chile, Santiago, Chile\\
$^4$Department of Geography and Earth Sciences, Aberystwyth University, Aberystwyth, UK\\
$^5$Geophysical Institute, University of Alaska Fairbanks, Fairbanks, AK, USA\\
$^6$Scott Polar Research Institute, University of Cambridge, Cambridge, UK\\
Correspondence: Dylan Kreynen, E-mail: \email{kreynen@uio.no}}

\begin{frontmatter}

\maketitle 

\begin{abstract}
Basal channels, troughs carved into ice shelf bases by buoyant plumes, modulate basal melt and can weaken ice shelves. Through hydrostatic adjustment they have a surface expression, yet their geometry is largely mapped manually, limiting the number of channels studied and the consistency between mappings. We present Sub-Ice, a semi-automated tool that maps basal channels from surface digital elevation models. From user-defined start and end points, Sub-Ice traces the channel centerline by iteratively locating elevation minima, extracts cross-sectional profiles and delineates channel edges using one of three methods (slope threshold, profile knee point or peak prominence), producing georeferenced centerlines and edges and along-channel width and depth. Applied to 34 channels on Roi Baudouin Ice Shelf, East Antarctica, Sub-Ice centerlines differ from manually digitized ones by a median of 132 m, at the upper end of the differences between manual records from different years (74-132 m), while following the channel trough more closely. Absolute channel widths depend on the edge method and the width of the edge search window, which we calibrate against a slope-based criterion. Rather than seeking a single ``true'' edge, Sub-Ice applies explicit, reproducible criteria, enabling consistent comparison of channel geometry across channels, ice shelves and time.
\end{abstract}

\end{frontmatter}

\section{Introduction}

Ice shelves (the floating extensions of an ice sheet) exert a backstress on the glaciers and ice streams that feed them, slowing the flow of ice toward the sea. A reduction in this buttressing effect due to ice shelf thinning or break-up can trigger an acceleration of upstream ice discharge and, in turn, an indirect contribution to sea level rise \citep{rignot_accelerated_2004, furst_safety_2016}. A key driver of ice shelf thinning is ocean-driven basal melting, and one of its most visible consequences at the ice shelf base is the formation of basal melt channels: elongated, kilometer-scale troughs carved by buoyant water plumes. This buoyant water can originate from subglacial discharge at the grounding line \citep{le_brocq_evidence_2013} or from ocean water intruding beneath the shelf and concentrating in pre-existing basal topography \citep{gladish_ice-shelf_2012, millgate_effect_2013}. Once established, channels tend to be self-sustaining and can persist over decades. Basal melt channels are widespread around the Antarctic continent and have been documented beneath ice shelves across a wide range of oceanographic and glaciological settings \citep{alley_impacts_2016}. Their relevance to ice shelf stability extends beyond their role as a signature of ocean forcing. Locally thinned ice above a channel concentrates strain \citep{lilien_effect_2025}, and channels have been linked to basal and surface crevassing \citep{rignot_channelized_2008, vaughan_subglacial_2012, alley_impacts_2016}, the weakening of ice shelf shear margins \citep{alley_troughs_2019, lhermitte_damage_2020, miele_marginal_2023}, and transverse fractures that can propagate across the shelf \citep{dow_basal_2018}. Channel-driven fracture has been directly associated with ice shelf calving events: for example, in 2016 the calving of iceberg C-33 from the Nansen Ice Shelf followed a transverse fracture that developed above a basal channel \citep{dow_basal_2018, dow_complex_2024}, and in 2021 iceberg D-29 calved from Roi Baudouin Ice Shelf following the joining of two channel-associated fractures \citep{jackson_channelised_2026}. Basal melt channels are, in this sense, ``both indicators of change and modulators of ice-shelf stability'' \citep{alley_role_2023}. Recently, it was shown that channelized melt beneath Antarctic ice shelves has been substantially underestimated, warranting further investigation into the link between basal channel dynamics and large-scale ice shelf stability \citep{zinck_channelized_2026}.

Studying basal channels is inherently difficult because they form at the ice-ocean interface, tens to hundreds of meters below the surface. In-situ measurements from ground-penetrating radar (GPR) surveys \citep{langley_complex_2014, drews_actively_2017}, boreholes \citep{stanton_channelized_2013} and autonomous underwater vehicles \citep{dutrieux_basal_2014, wahlin_swirls_2024} provide the most direct view of channel geometry and the ocean conditions within channels, but their coverage is limited in space and time. Numerical models offer insight into the plume dynamics and thermodynamics driving channel formation and evolution \citep{jenkins_convection-driven_2011, gladish_ice-shelf_2012, sergienko_basal_2013, millgate_effect_2013, drews_evolution_2015, stubblefield_linear_2023}, but are limited by the scarcity of observations at the ice-ocean interface. Remote sensing provides a complementary route to spatially distributed and temporally repeated observations. Although melting takes place at the ice shelf base, basal channels have a visible surface expression due to hydrostatic adjustment: because the ice above a channel is thinner than the surrounding shelf, the surface is slightly depressed, typically by only a few meters across channels several kilometers wide. This surface expression is a dampened reflection of the basal geometry, which can be considerably more complex \citep{langley_complex_2014}, and narrow channels can be held out of hydrostatic equilibrium by bridging stresses \citep{drews_evolution_2015}. Large channels, however, are close to hydrostatic equilibrium, and their surface expression therefore provides a useful first-order measure of channel position, shape and evolution.

Remote sensing studies have exploited both optical imagery, in which channels appear as subtle brightness contrasts, and digital elevation models (DEMs), in which they appear as topographic lows. \cite{alley_impacts_2016} compiled the first Antarctic-wide inventory of basal channel distribution, identifying channels manually in MODIS mosaics and Landsat 8 imagery. They quantified channel abundance by length and density (channel length per ice shelf area) rather than by channel area, as they found channel edges diffuse and difficult to identify consistently. Subsequent studies have focused on the temporal evolution of individual channels or channel systems. \cite{chartrand_basal_2020} traced the migration of a large channel on the Getz Ice Shelf by locating local minima in DEMs and altimetry profiles, which were then manually edited and smoothed into centerlines. \cite{alley_evolution_2024} manually digitized the centerlines of 12 large channels in West Antarctica over five decades of Landsat imagery, showing that ice-flow-independent lateral migration constitutes a novel proxy for changes in ocean forcing. Similarly, \cite{jackson_channelised_2026} manually mapped the centerlines of 11 channels on Roi Baudouin Ice Shelf annually from 2002 to 2021. Advances in altimetry now also allow channelized melt to be resolved directly: \cite{lowery_spatio-temporal_2025} derived monthly melt maps of the Pine Island Glacier ice shelf from CryoSat-2 DEMs, showing that melt varies both across and along its largest channels. Detailed channel geometries (centerlines, edges and cross-sectional shape) have thus been mapped for only a small number of channels, and a consistent record of basal channel geometry beyond individual ice shelves is still lacking. 

A first step toward automated mapping was taken by \cite{zinck_channelized_2026}, who applied watershed delineation to a downsampled surface DEM to extract channel centerlines across nine ice shelves. They also estimated channel width by fitting a Gaussian curve to elevation transects perpendicular to these centerlines. However, agreement with the manually mapped centerlines of \cite{alley_impacts_2016} varied considerably between ice shelves. In addition, width was represented by the standard deviation of the fitted curve rather than by the distance between channel edges, so channel edges and outlines were not delineated. Channel mapping (tracing centerlines, delineating edges and extracting cross-sectional profiles) has therefore largely remained a manual task. This is time-consuming, which limits the number of channels and the spatial and temporal coverage that can practically be studied. It also relies on the subjective judgment of individual analysts to decide where exactly a channel centerline or edge lies: in \cite{alley_evolution_2024}, independent repeat digitizations of the same channel differed through both random and interpretive errors, and comparisons had to be limited to records produced by the same investigator. As a result, channel geometries derived in different studies are difficult to compare, and results are hard to reproduce or extend to new data. 

Here we present Sub-Ice (\textbf{Su}rface-\textbf{b}ased \textbf{I}ce Shelf \textbf{C}hannel \textbf{E}xtraction; \citealp{kreynen_subice_2026}), a semi-automated MATLAB tool designed to address these limitations. Given a DEM and user-specified start and end points, Sub-Ice traces the channel centerline, extracts perpendicular cross-sectional profiles and delineates the channel edges, producing georeferenced shapefiles and quantitative channel geometry metrics. The tool can map multiple channels from a single DEM simultaneously, or track the evolution of a single channel across a time series of DEMs. We illustrate these capabilities through an application to Roi Baudouin Ice Shelf in East Antarctica, where we use Sub-Ice to map the 24 main and 10 tributary channels readily identifiable in the Reference Elevation Model of Antarctica mosaic (REMA; \citealp{howat_rema_2022}). Close agreement between automatically and manually delineated channel centerlines, together with explicit and reproducible edge detection criteria, highlights the potential of Sub-Ice to reliably and efficiently map basal channel geometry based solely on surface DEMs.

\section{Data and methods}

\subsection{Overview of Sub-Ice}

\begin{figure*}
\includegraphics[width=\textwidth]{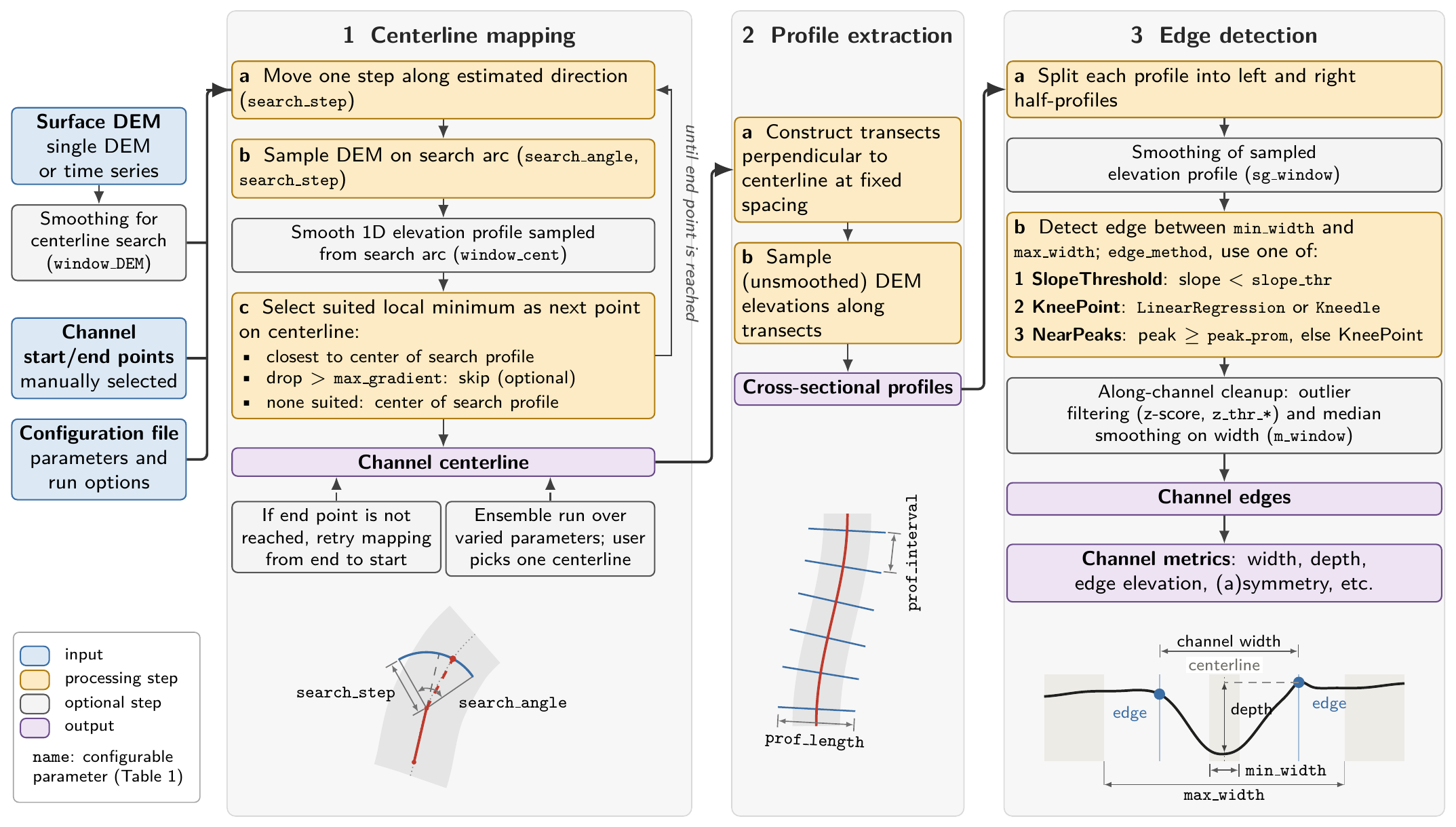}
\caption{Sub-Ice workflow. From a surface DEM and user-selected start and end points, Sub-Ice (1) maps the channel centerline, (2) extracts cross-sectional elevation profiles perpendicular to it and (3) detects the channel edges, from which channel metrics are derived. Purple boxes show outputs, each of which serves as input to the next step; gray boxes show optional steps. Parameters are set in the configuration file (Table \ref{tab:parameters}). Sketches illustrate each stage. Outputs are saved as tables and georeferenced shapefiles, and visualized in map-view, across-channel and along-channel figures. }
\label{flowchart_methods}
\end{figure*}

Sub-Ice is a semi-automated MATLAB algorithm for mapping the centerline and edges of ice shelf basal melt channels based on their surface expression in a DEM. Given a DEM and user-defined start and end points for each channel of interest, the algorithm (1) maps the channel centerline, (2) extracts perpendicular cross-sectional elevation profiles and (3) delineates the channel edges. From these profiles, channel characteristics including channel width, depth and edge elevation are derived, and outputs are saved as tables, figures and georeferenced shapefiles. The algorithm can be run in two modes: mapping one or more channels from a single DEM, or tracking a single channel across a time series of DEMs to study its temporal evolution. The main workflow is outlined in Fig. \ref{flowchart_methods}, and individual steps are detailed below. The code is openly available at \url{https://github.com/DylanKreynen/Sub-Ice}. 

\subsubsection{Centerline mapping}

Sub-Ice reconstructs a channel's centerline (the line connecting the channel's topographic lows) between user-defined start and end points. Working along the estimated channel direction, the algorithm samples the DEM at regular intervals, identifies suitable local minima and connects them into a centerline (sketch in Fig. \ref{flowchart_methods}; parameters in Table \ref{tab:parameters}). The following steps are repeated until the end point is reached: 

\begin{enumerate}[noitemsep]
    \item Step a fixed distance (\texttt{search\_step} [m]) from the last centerline point in the estimated along-channel direction. 
    \item Construct a circular arc of sampling points spanning an angular window \texttt{search\_angle} [$^\circ$]. 
    \item Sample the DEM along this arc using bilinear interpolation, optionally smoothing the resulting profile with a moving mean filter (window \texttt{window\_cent} [m]). 
    \item Select the next centerline point from the local minima on the arc: 
        \begin{itemize}[noitemsep]
        \item[$\cdot$] Minima for which the centerline gradient from the last centerline point exceeds \texttt{max\_gradient} [\%] are rejected. 
        \item[$\cdot$] Of the remaining minima, the one that best preserves the direction of the last centerline segment is selected. 
        \item[$\cdot$] If no suitable minimum remains, the center of the arc is selected.
        \end{itemize}
\end{enumerate}

The centerline search is thus direction-preserving: it always looks for a topographic low in the established centerline direction (or toward the end point, for the first step). Combined with optional DEM smoothing (2D moving mean, window \texttt{window\_DEM} [m]), this reduces the risk of the algorithm being deflected into other surface lows such as crevasses (which often run perpendicular to the channel), DEM noise or artifacts. 

The search stops once the latest centerline point lies within one \texttt{search\_step} of the end point, which is then appended to the centerline. To prevent runaway searches, the search is aborted once the centerline length exceeds twice the straight-line distance between start and end points. This factor is large enough to allow meandering centerlines that are still on track, but small enough to stop failed searches early. If the search is aborted, the algorithm can optionally retry with start and end points reversed and, if still unsuccessful, with randomly perturbed search parameters, until the end point is reached or the maximum number of attempts (\texttt{max\_recursions}) is reached. Perturbed values of \texttt{search\_step}, \texttt{search\_angle}, \texttt{max\_gradient} and \texttt{window\_cent} are drawn independently from each other for each attempt from a normal distribution centered on the configured value, with a standard deviation of 5\%. The parameter values of the successful attempt, and whether start and end points were reversed, are saved as attributes in the output centerline shapefile. 

As the final centerline depends on the search parameters and on the surface characteristics of the ice shelf and channel, a single parameter set does not always produce an appropriate centerline. In such cases, we recommend running the centerline search in ensemble mode. The user specifies a set of values for any or all of the centerline search parameters (Table \ref{tab:parameters}), and Sub-Ice maps the centerline for every combination of these values (a full-factorial ensemble). Realizations that fail to connect the start and end points are discarded, and the user selects a representative centerline from the remainder by visual inspection. This centerline is then used for profile extraction and edge detection.

\subsubsection{Profile extraction}

Cross-sectional elevation profiles are extracted perpendicular to the centerline at regular intervals. By default, one profile is placed at the midpoint of each centerline segment (the straight line connecting two consecutive centerline points), so that profile spacing equals \texttt{search\_step}. Alternatively, profiles can be placed at a fixed along-channel spacing \texttt{prof\_interval} [m], independent of the search step. Each profile is oriented perpendicular to its centerline segment and sampled over a length \texttt{prof\_length} [m] at DEM resolution using bilinear interpolation, with sampling points distributed symmetrically so that the central point coincides with the centerline. Profiles are always sampled from the original, non-smoothed DEM. All profiles are oriented consistently: their left and right halves correspond to the left- and right-hand sides of the channel looking downstream (from start to end point), a convention used for all profile and edge output.

\subsubsection{Edge detection}

\begin{figure*}
\centering
\includegraphics[width=0.9\textwidth]{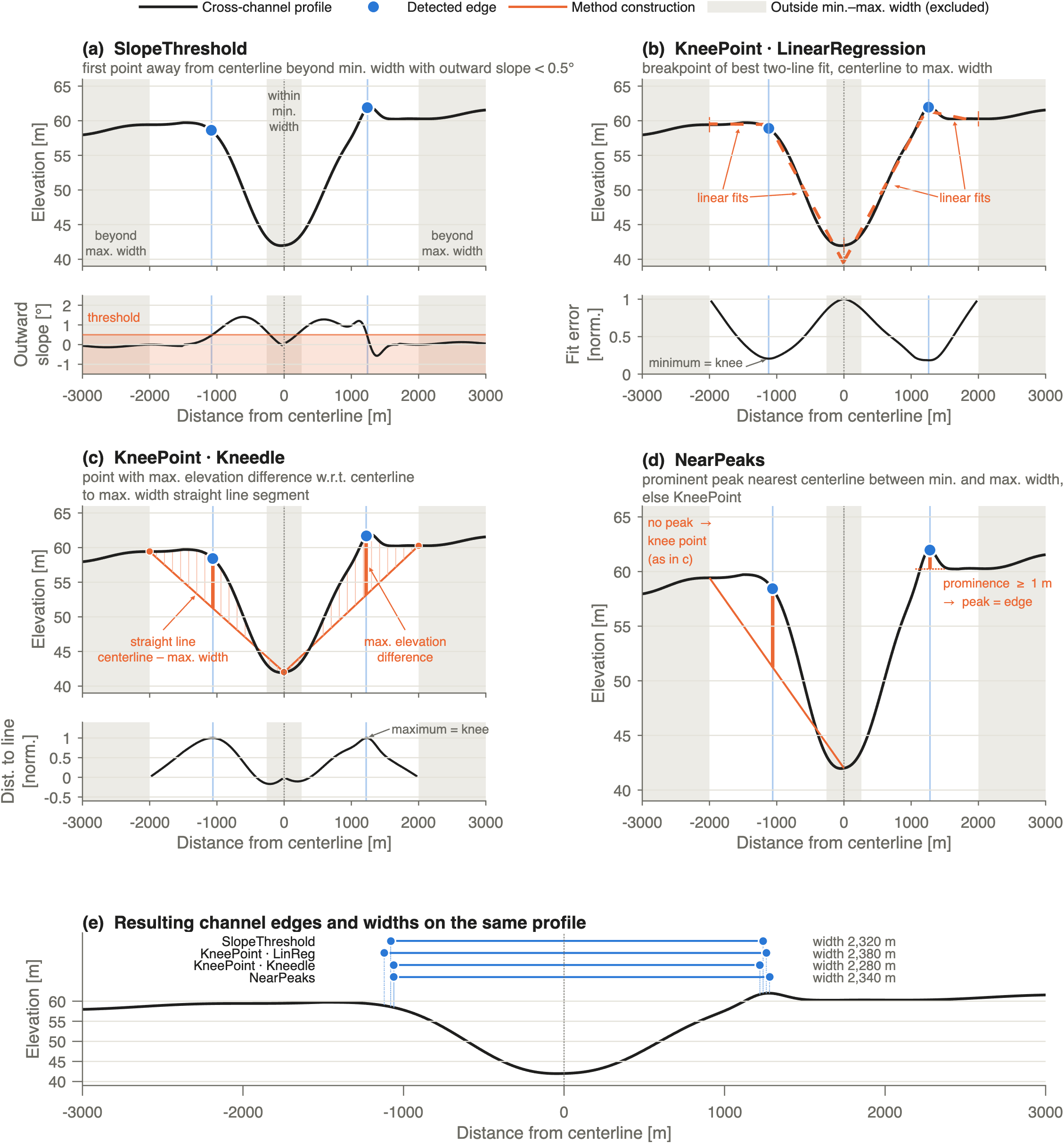}
\caption{Channel edge detection methods applied to a synthetic cross-sectional channel profile. Edges (blue dots) are identified on both sides of the centerline between the minimum and maximum width (gray shading marks the excluded zones). (a) SlopeThreshold: first point where the outward slope drops below a threshold. (b) KneePoint-LinearRegression: two lines are fitted to the half-profile from the centerline to the maximum width; the edge is the breakpoint that minimizes the combined absolute error of both fits. (c) KneePoint-Kneedle: point farthest above the straight line from the centerline to the maximum width. (d) NearPeaks: nearest prominent peak (otherwise KneePoint-Kneedle). (e) Edges and channel widths from all three methods including both KneePoint variants. Lower plots in (a)–(c) show the quantity each method evaluates.}
\label{edge_method_concepts}
\end{figure*}

Channel edges, the locations where the channel flank transitions to the surrounding ice shelf surface, are identified automatically on each cross-sectional profile. All implemented edge detection methods operate on each half-profile (left and right of the centerline) separately, searching outward from the centerline. The search is limited to a user-defined width range: edges are placed no closer to the centerline than half of \texttt{min\_width} [m] and no farther than half of \texttt{max\_width} [m], with the profile beyond \texttt{max\_width} being ignored (separate limits can be set for the left and right side if desired). Prior to edge detection, profiles can be smoothed using a Savitzky-Golay filter (polynomial degree 3, window size \texttt{sg\_window} [m]); final edge elevations are always taken from the non-smoothed profile. The user selects one of three edge detection methods (Fig. \ref{edge_method_concepts}):

\begin{enumerate}[noitemsep]
    \item \textit{Slope threshold method} (\texttt{SlopeThreshold}): the edge is placed at the first point beyond the minimum width where the outward surface slope falls below a user-defined threshold angle (\texttt{slope\_thr} [$^\circ$], default 0$^\circ$). If the threshold is not reached, the edge is placed at the maximum width. 
    \item \textit{Knee point method} (\texttt{KneePoint}): the edge is placed at the knee of the half-profile, i.e., the point where the elevation transitions from a rapid rate of change along the channel flank to a more gradual one beyond it. Two algorithms are implemented to locate the knee (\texttt{knee\_method}): 
        \begin{itemize}[noitemsep]
        \item[$\cdot$] \texttt{LinearRegression} moves outward along the half-profile one point at a time, fits a straight line on either side of that point, and selects the point that minimizes the combined absolute error of both fits. 
        \item[$\cdot$] \texttt{Kneedle} draws a straight line from the centerline point to the outer end of the half-profile, and selects the point with the largest elevation above this line. 
        \end{itemize}
    If the knee lies within the minimum width, the edge is moved out to the minimum width. 
    \item \textit{Near peaks method} (\texttt{NearPeaks}): the edge is placed at the topographic peak closest to the centerline, considering only peaks with a minimum prominence (\texttt{peak\_prom} [m]) between the minimum and maximum width. If no such peak is found, the method falls back to the knee point method. 
\end{enumerate}

When concatenated along the channel, the left and right edges of all profiles form the left and right channel outlines. As edges are detected on each profile independently, raw outlines can be noisy due to outliers on individual profiles. Sub-Ice offers two optional ways to address this. First, outlines can be smoothed along the channel by applying a median filter to edge positions (window size \texttt{m\_window}, in number of profiles); with the near peaks method, detected peaks can be excluded from smoothing (\texttt{keep\_pks}). Second, outliers can be identified based on the z-score of edge positions (\texttt{z\_thr\_idx}) and edge elevations (\texttt{z\_thr\_elev}) along the channel, with separate thresholds for the left and right edges. Flagged edges are either omitted or replaced by the moving median of neighboring edges (window size \texttt{edge\_subst\_window}, in number of profiles). As smoothing is applied before outlier detection, we recommend using either approach, but not both. 

The channel trough is taken to be the central point of each profile, i.e., the centerline position. As the centerline may be traced on a smoothed DEM while profiles are sampled from the original DEM, trough elevations can be affected by noise or artifacts. Trough outliers are therefore flagged based on their z-score along the channel (\texttt{z\_thr\_trough\_elev}) and either omitted or replaced by the moving mean of neighboring troughs (window \texttt{trough\_subst\_window}, in profiles). Sub-Ice then derives the channel width (distance between the edges), the left and right half-width (distance of each edge from the centerline) and the channel depth (trough elevation below the left edge). Depth is defined relative to the left edge because, looking downstream, this side of Antarctic basal channels is generally steeper and more distinct due to Coriolis-enhanced melt \citep{sergienko_basal_2013, gourmelen_channelized_2017, alley_evolution_2024}. 

\subsubsection{Setup, input and output}

A Sub-Ice run is set up through a configuration file, in which the user specifies the input files, the parameters for centerline mapping, profile extraction and edge detection (Table \ref{tab:parameters}), and the output settings (which figures, tables and shapefiles to save, where, and at what resolution). Configuration files can be saved for reuse and allow Sub-Ice to be run in batches. 

The required input is a surface DEM in GeoTIFF format with a projected coordinate system (map units in meters); no-data values are set to NaN upon loading. If DEM smoothing is requested for centerline mapping (\texttt{window\_DEM} [m]), the smoothed DEM is cached to disk for reuse in subsequent runs. Channel start and end points, which define the upstream and downstream extent of each channel, are provided in one of three ways (\texttt{start\_end\_method}): interactively, by clicking on the DEM in a MATLAB figure window; from a multipoint shapefile in the same projection as the DEM; or by entering image coordinates directly. The shapefile option is preferred for reproducibility. 

Centerlines, left and right edges and profile transects are written as georeferenced ESRI shapefiles in the projection of the input DEM. Cross-sectional profile elevations, channel edge locations and elevations, several channel metrics (width, depth) and related statistics are saved as CSV tables for easy further analysis. Output figures include map views of the mapped channels, cross-sectional profiles (full and limited to the channel edges) colored by along-channel position, and along-channel plots of edge and trough elevation, channel depth, channel width and mean surface elevation.

\begin{table*}
\centering
\caption{User-definable Sub-Ice parameters, set in the configuration file, with the values used for the Roi Baudouin Ice Shelf case study. Centerlines were mapped with an ensemble approach: parameters given as sets \{\ldots\} were combined in a full-factorial ensemble of $3 \times 10 \times 2 \times 4 = 240$ centerline realizations per channel. Only realizations connecting a channel's start and end points were retained, and one centerline per channel was selected by visual inspection for further analysis. ST = SlopeThreshold, KP = KneePoint, NP = NearPeaks, thr. = threshold. }
\label{tab:parameters}
\footnotesize
\setlength{\tabcolsep}{4pt}
\renewcommand{\arraystretch}{0.95}
\begin{tabular}{l l p{8cm} c}
\hline
\textbf{Group} & \textbf{Parameter} & \textbf{Description} & \textbf{Value} \\
\hline
Start/end & \texttt{start\_end\_method} & method (1 click, 2 shapefile, 3 manual) & 2 \\
\hline
\multirow{6}{*}{Centerline}
 & \texttt{window\_DEM} & DEM smoothing window [m] (0 = none) & \{100, 250, 500\} \\
 & \texttt{search\_step} & step distance for centerline search [m] & 1000 \\
 & \texttt{search\_angle} & search arc angular width [$^\circ$] & \{30, 40, \ldots, 120\} \\
 & \texttt{max\_gradient} & max. elevation drop from last centerline point [\%] & \{2, 100\} \\
 & \texttt{window\_cent} & search profile smoothing window [m] & \{0, 50, 100, 200\} \\
 & \texttt{max\_recursions} & max. no. of search attempts [-] \newline (1 = no retries, 2 = reversal only) & 500 \\
\hline
\multirow{2}{*}{Profiles}
 & \texttt{prof\_length} & cross-section profile length [m] & 12000 \\
 & \texttt{prof\_interval} & profile spacing along centerline [m] (0 = per segment) & 500 \\
\hline
\multirow{9}{*}{Edges}
 & \texttt{edge\_method} & edge method (ST/KP/NP) & KneePoint (KP) \\
 & \texttt{knee\_method} & knee point method (LinearRegression/Kneedle) & Kneedle \\
 & \texttt{min\_width} & minimum channel width [m] (0 = none) & 1000 \\
 & \texttt{max\_width} & maximum channel width [m] (0 = \texttt{prof\_length}) & 8000 \\
 & \texttt{sg\_window} & profile smoothing window, Savitzky-Golay [m] & 500 \\
 & \texttt{m\_window} & edge smoothing window, median filter [-] & 0 \\
 & \texttt{slope\_thr} & edge slope threshold [$^\circ$] (only ST) & 0.00 \\
 & \texttt{peak\_prom} & minimum peak prominence [m] (only NP) & 0.25 \\
 & \texttt{keep\_pks} & keep peaks fixed during smoothing [bool] (only NP) & 0 \\
\hline
\multirow{5}{*}{Outliers}
 & \texttt{z\_thr\_elev} & edge elevation outlier thr., z-score [L, R] (0 = none) & [0, 0] \\
 & \texttt{z\_thr\_idx} & edge position outlier thr., z-score [L, R] (0 = none) & [1.75, 1.75] \\
 & \texttt{edge\_subst\_window} & edge outlier substitution window [-] & 15 \\
 & \texttt{z\_thr\_trough\_elev} & trough elevation outlier thr., z-score (0 = none) & 0 \\
 & \texttt{trough\_subst\_window} & trough outlier substitution window [-] & 3 \\
\hline
\end{tabular}
\end{table*}

\subsection{Application to Roi Baudouin Ice Shelf}

We demonstrate Sub-Ice on Roi Baudouin Ice Shelf, where we (i) map all channels readily identifiable in a single DEM mosaic; (ii) evaluate the resulting centerlines against an independent record of manually digitized centerlines; and (iii) assess the sensitivity of the mapped channel edges to the edge detection settings.

\subsubsection{Study site}

Roi Baudouin Ice Shelf is located in Dronning Maud Land, East Antarctica (Fig. \ref{overview_shelves}). Fed by the West and East Ragnhild Glaciers, it covers $\sim$33,000 km$^{2}$ and extends $\sim$100 km into the Riiser-Larsen Sea. The Derwael Ice Rise and several smaller pinning points near the ice front provide substantial buttressing \citep{berger_control_2016}. The ice shelf has a cold cavity, with seasonally varying basal melt rates of up to 10 m a$^{-1}$ \citep{sun_topographic_2019}. Much of this melt is channelized: the ice shelf, particularly in its western part, hosts a well-developed network of basal channels that are clearly expressed at the surface in both optical imagery and DEMs. Most channels mapped to date originate at subglacial meltwater outlets along the grounding line \citep{drews_actively_2017, drews_atmospheric_2020}, while ocean intrusions modulate the heat supplied to the cavity \citep{zhou_winddriven_2014, sun_topographic_2019}. Several channels have been studied in detail with GPR and surface observations \citep{drews_evolution_2015, drews_actively_2017}, and their activity has varied over decadal timescales \citep{zhou_reactivation_2025}. Channelized melt has also weakened the western ice shelf, with channel-associated fracture preceding the calving of iceberg D-29 in 2021 \citep{jackson_channelised_2026}. Its dense, well-studied channel network and an independent record of manually digitized centerlines make Roi Baudouin Ice Shelf a suitable test case for Sub-Ice.

\subsubsection{Data}

Surface elevation data come from the Reference Elevation Model of Antarctica (REMA), which offers Antarctic-wide coverage at a spatial resolution of up to 2 m and is freely available. REMA is constructed by stereophotogrammetry from Maxar satellite imagery acquired over austral summers between 2009 and 2021 (version 2 mosaic), with each strip DEM vertically registered to satellite altimetry measurements \citep{howat_rema_2022}. Here we use the mosaic at 10 m spatial resolution. Each mosaic pixel represents the median elevation of all strips available at that location. The number and timing of contributing observations therefore vary from pixel to pixel, and no single acquisition date can be assigned to the mosaic. Tiles were downloaded individually and merged to cover Roi Baudouin Ice Shelf.

Channels were identified visually from the REMA mosaic by one experienced investigator, who had not seen the manually digitized centerlines of \cite{jackson_channelised_2026}. We define a basal channel here as an elongated surface depression that can be traced continuously over at least 10 km (including the main channel in case of tributary channels). In total, 24 main channels and 10 tributaries (channels that merge into a main channel) were identified. Start and end points were placed at the most upstream and downstream locations where each channel could be traced continuously and saved as a multipoint shapefile for reproducibility. Channel M8 was mapped in two segments (M8 and M8b; Fig. \ref{overview_shelves}) because it is interrupted by a large rift. The two segments were merged after edge detection and are treated as a single channel.

As a reference, we use the manually digitized channel centerlines of \cite{jackson_channelised_2026}, who mapped 11 channels on Roi Baudouin Ice Shelf for each year from 2002 through 2021. They used Landsat 7 ETM+ and Landsat 8 OLI scenes acquired between November and February with solar elevations above 20$^\circ$ to manually delineate channel centerlines at $\sim$300 m intervals, guided by the light-shadow contrast across each channel's surface depression, which they enhanced by adjusting the brightness and contrast of individual scenes. Their annual repeat digitizations show that several of the channels have a persistent, spatially fixed source that produces a continuous surface trace, which is advected with ice flow. As a result, channel position varies little from year to year along sections aligned with ice flow, which make up most of the channel length (Fig. S6). Hereafter, we refer to these as the manual centerlines, and to the centerlines mapped with Sub-Ice as the Sub-Ice centerlines. We compare the Sub-Ice centerlines against the 2018 manual centerlines, which include all 11 channels. Although the REMA mosaic strips span 2009-2021, over Roi Baudouin Ice Shelf most were acquired in the second half of this period, with the largest number in 2018 (REMA tile metadata). The 2015 and 2021 manual centerlines (the nearest years before and after 2018 in which all 11 centerlines were digitized) are used to quantify the year-to-year variability of the manual record itself (Fig. S6).

\subsubsection{Sub-Ice configuration}

Sub-Ice configuration and parameter values are listed in Table \ref{tab:parameters}. Centerlines were mapped in ensemble mode, combining three DEM smoothing windows, 10 search angles, two gradient thresholds (2\% and 100\%, the latter of which effectively disables the criterion) and four search profile smoothing windows into 240 realizations per channel. Realizations that did not connect start and end points were discarded. Without search parameter perturbation but allowing the search to retry with start and end points reversed (\texttt{max\_recursions = 2}), 50\% (main channels) and 56\% (tributaries) of realizations connected their respective start and end points. Success rates were most sensitive to \texttt{search\_angle}, and largely insensitive to the chosen smoothing windows (\texttt{window\_DEM}, \texttt{window\_cent}) and gradient threshold (\texttt{max\_gradient}). Allowing for parameter perturbation (\texttt{max\_recursions} = 500) increased success rates to 99\% and 97\%, respectively. From this last ensemble, one centerline per channel was selected for further analysis by visual inspection against the REMA mosaic (see Figs. S1-S2), without having seen the manual centerline record. 

Channel edges were detected with the \texttt{KneePoint} method using the \texttt{Kneedle} knee point algorithm, on 12 km long cross-sectional profiles placed every 500 m along the centerline. Edge outliers (z-score of the edge position above 1.75) were replaced by the moving median over 15 profiles (7.5 km). Unless stated otherwise, all results presented in this study are based on this configuration. We set \texttt{max\_width} to 8 km, where the mean width of the KneePoint methods matches the stable mean width of SlopeThreshold (Section \ref{results_edges}).

\subsubsection{Evaluation}
\label{evaluation}

\begin{figure*}
\centering
\includegraphics[width=0.94\textwidth]{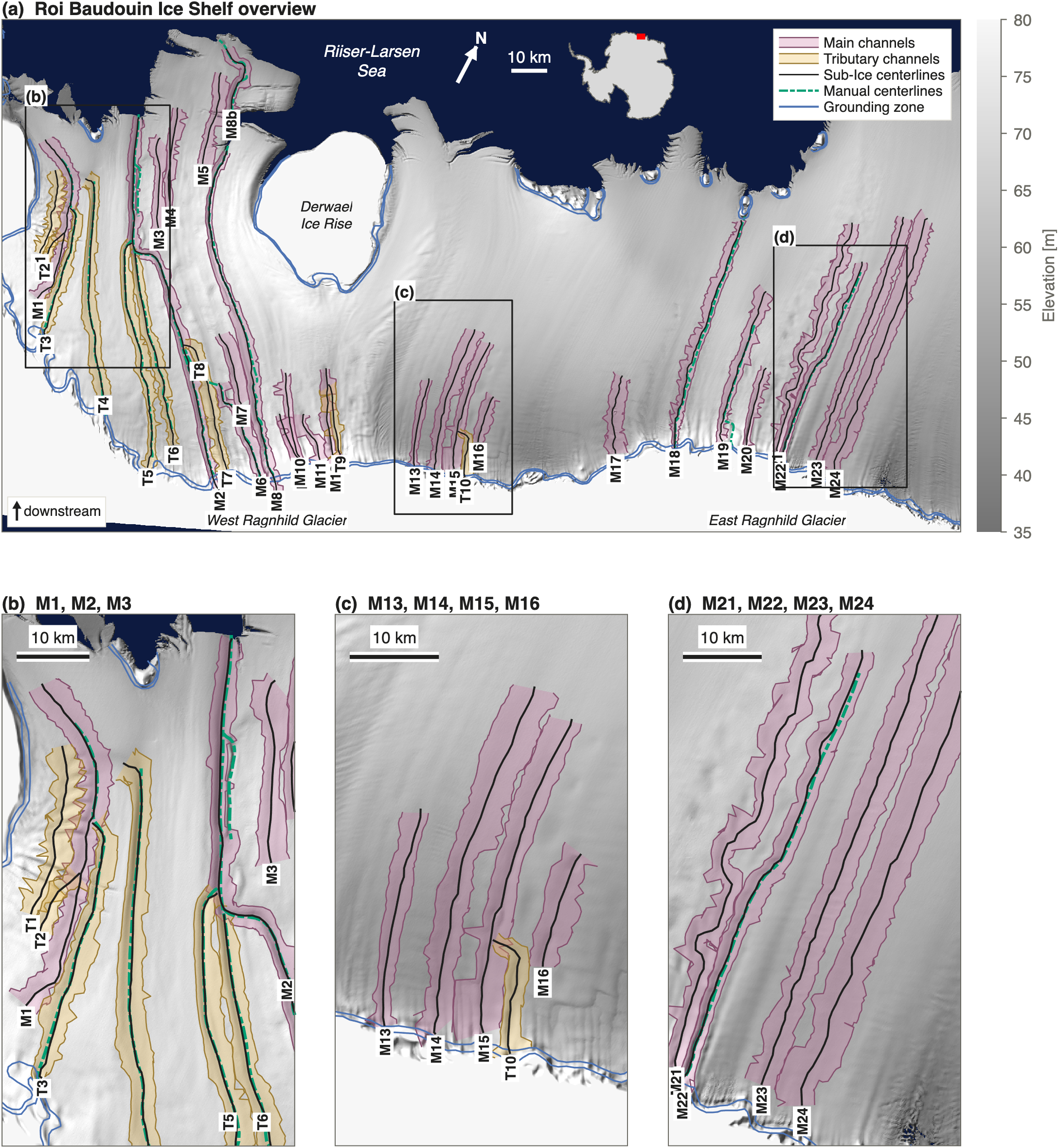}
\caption{Channel centerlines and edges mapped by Sub-Ice on Roi Baudouin Ice Shelf, East Antarctica, on hillshaded REMA surface elevation \citep{howat_rema_2022}. (a) Overview of all 34 channels (24 main channels, M1–M24, with M8 mapped in two parts, M8 and M8b; 10 tributaries, T1–T10); the red square in the inset shows the location in Antarctica. Centerlines are black and labeled at their upstream end as in Figs. \ref{example_across} and \ref{example_along}; the area between the channel edges is shaded purple for main channels and orange for tributaries. Edges were detected on cross sections every 500 m along each centerline with KneePoint-Kneedle (Table \ref{tab:parameters}). Dash-dotted green: 2018 manual centerlines \citep{jackson_channelised_2026}. Grounding zones are from MEaSUREs version 1 \citep{rignot_measures_2022}. (b–d) Close-ups of the areas outlined in (a): (b) M1–M3 and T1–T6, (c) M13–M16 and T10, (d) M21–M24. Per-channel maps, profiles and along-channel elevation, width and depth are shown in Figs. S7–S41.}
\label{overview_shelves}
\end{figure*}

We evaluate Sub-Ice centerlines against the 2018 manual centerlines, quantifying their agreement using the cross-sectional profiles that Sub-Ice constructs perpendicular to its own centerline. For each profile, we locate the point where it intersects a manual centerline and define the local offset as the horizontal distance between this intersection and the profile center. We also record the side on which the manual centerline lies (left or right, looking downstream), which gives a signed offset. Because each 12 km profile may also intersect the manual centerline of a neighboring channel, only intersections between the channel's mapped edges are retained. Because the two datasets divide the channel network differently (e.g., the manual centerline of M2 continues along what we identified as tributary T6, Fig. \ref{overview_shelves}), channels are not paired by label: each Sub-Ice profile is evaluated against whichever manual centerline crosses it. To test whether results depend on using the Sub-Ice centerline as the reference frame, we also repeat the comparison in reverse, constructing profiles perpendicular to the manual centerlines, detecting channel edges on these profiles with the same settings, and retaining only Sub-Ice centerline crossings that lie between these edges. Finally, applying the same procedure to pairs of manual centerlines from different years (2015, 2018 and 2021) provides an estimate of the offsets that can be expected from channel migration and digitization uncertainty alone.

Channel edges, in contrast, have not previously been mapped systematically for ice shelf basal channels, so no independent dataset is available against which to evaluate Sub-Ice's edge detection. Instead, we evaluate the sensitivity of the mapped edges to the choice of edge detection method and its parameters. We first perform a one-at-a-time sensitivity analysis, in which a single edge detection parameter is varied while all others are kept at the default values used in this study (Table \ref{tab:parameters}). The analysis covers SlopeThreshold, KneePoint (with Kneedle and LinearRegression methods), and NearPeaks (with KneePoint-Kneedle as fallback). The parameters varied are the maximum channel width (\texttt{max\_width}, 2-12 km, with \texttt{min\_width} fixed at 1 km), the profile smoothing window (\texttt{sg\_window}, 0-2 km), the minimum channel width (\texttt{min\_width}, 0–5 km), the slope threshold of SlopeThreshold (\texttt{slope\_thr}, at $-0.01^\circ$, $0.00^\circ$ and $0.01^\circ$) and the minimum peak prominence of NearPeaks (\texttt{peak\_prom}, at 0.25, 0.5 and 1.0 m). For each run, we compute the mean channel width (the mean of the per-channel mean widths), the change in width for each individual channel, and the fraction of edges placed at a search limit (i.e., where no edge was found between \texttt{min\_width} and \texttt{max\_width}). Results for \texttt{max\_width}, which we use to set the default search window, are presented in Section \ref{results_edges}; results for the other parameters are provided in the Supplement (Figs. S3-S5). With \texttt{max\_width} fixed, we then compare the edges found by the alternative methods on individual channels and on a selection of cross-sectional profiles (Fig. \ref{edge_comparison}).

\section{Results and evaluation}

\begin{figure*}
\includegraphics[width=\textwidth]{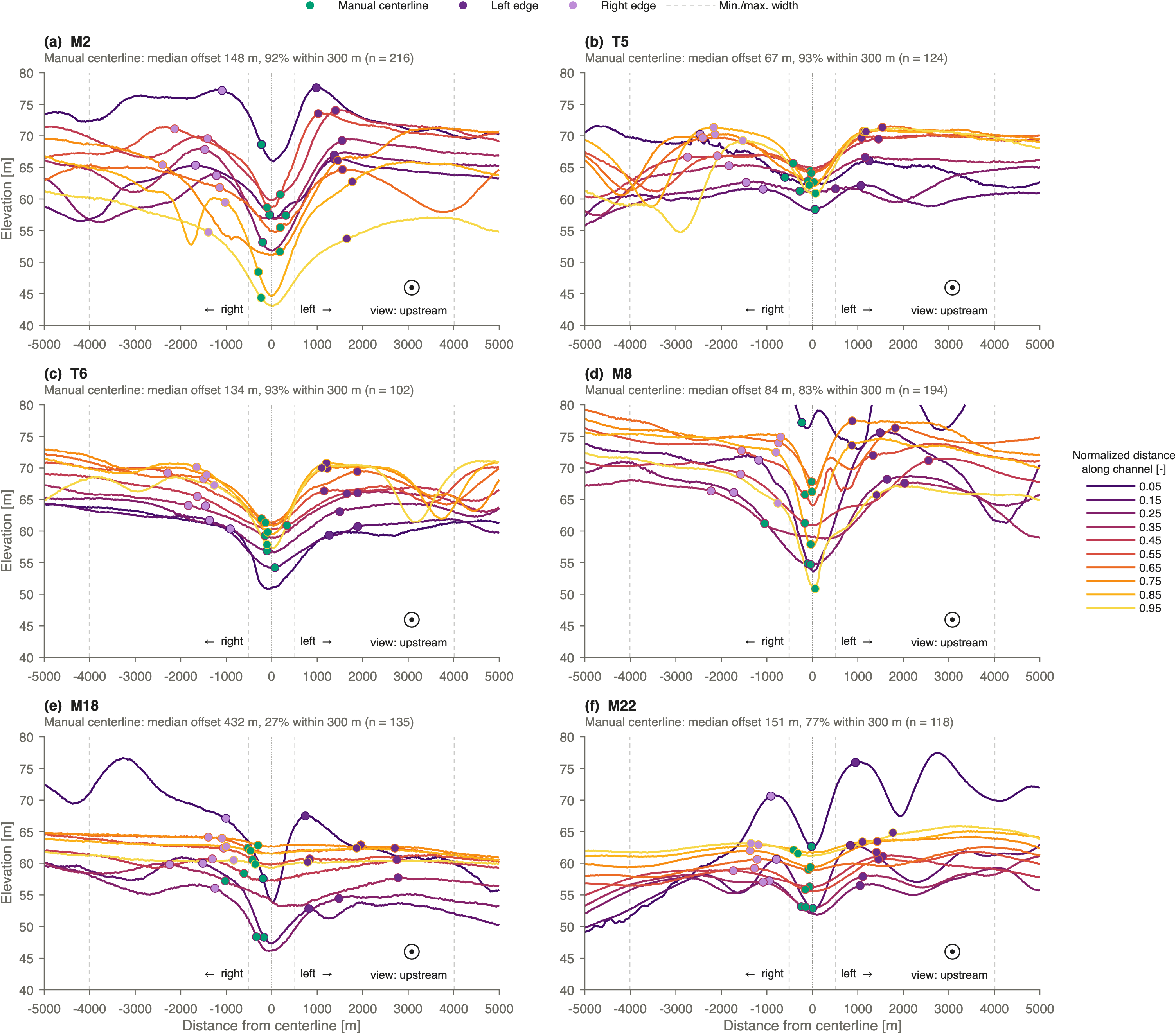}
\caption{Cross-sectional REMA elevation profiles across six example channels (all channels in Figs. S7–S41) on Roi Baudouin Ice Shelf, perpendicular to the Sub-Ice centerline (dotted line at 0 m). Ten profiles, at 0.05, 0.15, ..., 0.95 normalized distance along the channel, are highlighted in color (dark purple upstream to yellow downstream). Dark and light purple markers show the detected left and right channel edges on the highlighted profiles (defined looking downstream, KneePoint-Kneedle), with dashed lines at the edge search limits (1000 and 8000 m channel width). Green markers show where these profiles cross the 2018 manual centerline of \cite{jackson_channelised_2026}. Panel subtitles give the median offset between the two centerlines and the share of offsets within 300 m, over all $n$ cross sections where the manual centerline lies between the channel edges. Where the centerlines diverge, the Sub-Ice centerline tends to follow the trough more closely. The largest offset among well-sampled channels occurs along M18 (e). Channel labels correspond to those in Figs. \ref{overview_shelves} and \ref{example_along}.}
\label{example_across}
\end{figure*}

\begin{figure*}
\centering
\includegraphics[width=0.95\textwidth]{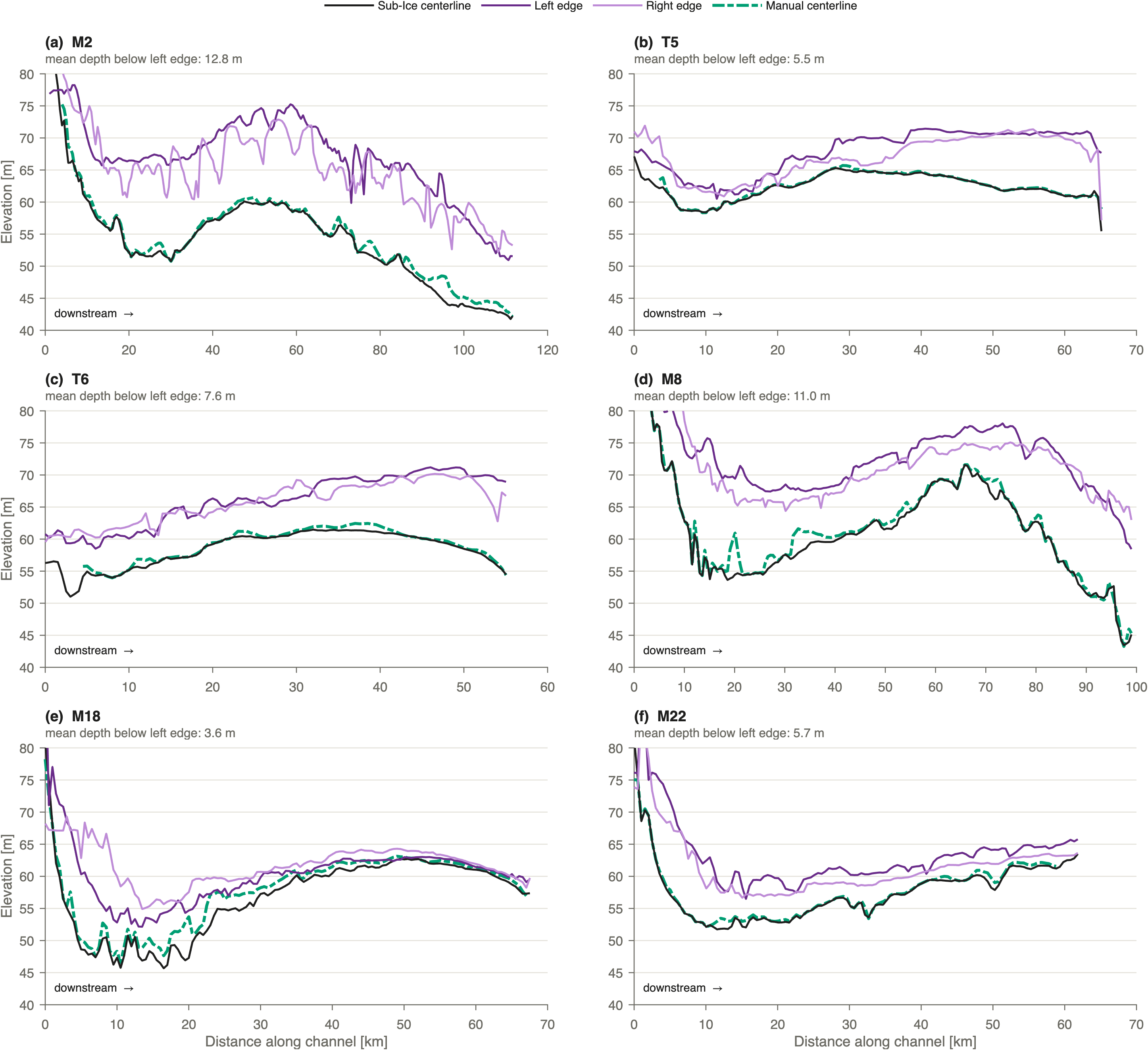}
\caption{Surface elevation along six example channels (all channels in Figs. S7–S41) on Roi Baudouin Ice Shelf, extracted from the REMA mosaic, plotted against distance along centerline from the channel start point. Black: elevation at the Sub-Ice centerline; dark and light purple: elevation at the left and right channel edges (defined looking downstream; KneePoint-Kneedle); dash-dotted green: elevation at the 2018 manual centerline of \cite{jackson_channelised_2026}. Panel subtitles give the mean channel depth, measured as the height of the left edge above the Sub-Ice centerline. Channel labels correspond to those in Figs. \ref{overview_shelves} and \ref{example_across}.}
\label{example_along}
\end{figure*}

\subsection{Centerlines and comparison with manual mapping}

In total, we mapped 34 channels on Roi Baudouin Ice Shelf: 24 main channels and 10 tributaries, with a combined centerline length of $\sim$1,380 km (Fig. \ref{overview_shelves}). Centerline lengths range from 7 km (T8) to 120 km (M8 including M8b), with a mean of 41 ± 28 km (mean ± 1 standard deviation). Main channels are on average longer than tributaries (44 ± 30 km and 33 ± 22 km, respectively).

The Sub-Ice and manual centerlines agree well (Fig. \ref{overview_shelves}; per-channel maps in Figs. S7-S41). All manual centerlines have a Sub-Ice counterpart, apart from a short (15 km) tributary of M2 that we did not identify as a separate channel, and only at the start of M19 do the two datasets follow different branches. Of the 1411 Sub-Ice profiles crossed by a manual centerline, 1288 (91\%) are crossed between the Sub-Ice channel edges and are used below. Most of the remaining 123 crossings belong to the manual centerline of a neighboring, parallel channel, and are counted from that channel's perspective. The median offset is 132 m (interquartile range 57–250 m), with 81\% of offsets within 300 m and 94\% within 500 m, a small fraction of the mapped channel widths. Per channel (Table S2), the median offset ranges from 67 to 176 m, except for M18 (432 m) and T7 (452 m, from only 16 cross sections). Reversing the roles of the two datasets (constructing profiles perpendicular to the manual centerlines and looking for Sub-Ice centerline crossings) gives nearly identical results (median 135 m, 80\% within 300 m), so the offsets are not an artifact of the profile construction. For comparison, manual centerlines from different years differ by a median of 74 m (2015–2018), 96 m (2018–2021) and 132 m (2015–2021), and the Sub-Ice centerlines agree more closely with the 2018 manual centerlines than with those of 2015 (152 m) or 2021 (169 m; Table S3). 

Where the two centerlines diverge, the Sub-Ice centerline follows the channel trough more closely (Figs. \ref{example_across} and \ref{example_along}). In 84\% of cross sections the surface elevation at the manual centerline is higher than at the Sub-Ice centerline (median difference 0.24 m), rising to 96\% for offsets larger than 100 m (median 0.53 m). This is partly by design: Sub-Ice traces the elevation minimum in the DEM, whereas the manual centerlines were digitized from contrast in coarser-resolution optical imagery. When the goal is to map the channel trough, Sub-Ice therefore offers an accuracy advantage over manual digitization from imagery alone, even where both agree on the channel's overall course.

Offsets are largest where channels are shallow. In the shallowest third of cross sections (depth $<$ 5.7 m), the median offset is 188 m and 65\% of offsets lie within 300 m, compared with 100 m and 92\% in the deepest third (depth $>$ 10.4 m). Conversely, cross sections with offsets larger than 300 m have a median depth of 4.1 m, half that of the remaining cross sections (8.6 m). This pattern is most pronounced along M18 (median offset 432 m), one of the shallowest channels compared, but offset also decreases with depth within 9 of the 12 channels. 

Unlike the year-to-year offsets within the manual record, the offsets between Sub-Ice and manual centerlines have a consistent direction: the manual centerline lies to the right of the Sub-Ice centerline (looking downstream) in 74\% of cross sections. This is most pronounced along M18, where it lies on the channel's right flank for all 135 cross sections (Figs. \ref{example_across} and S25). Neither channel migration nor random digitization error explains this pattern. The rightward displacement is present in each manual year (70\%, 74\% and 75\% of cross sections for 2015, 2018 and 2021), whereas the shifts between manual years are smaller and less consistent: from 2015 to 2018 the manual centerlines moved left and right equally often (median 1 m), and only from 2018 to 2021 did they shift to the right (median 47 m), most strongly along M1, M6, M19 and T3. The manual centerlines of M8 and M18 show no net lateral shift between 2015 and 2021 (signed median within 25 m), yet lie 80–430 m to the right of the Sub-Ice centerline. Random errors, in turn, would largely cancel out over 1288 cross sections. The displacement is consistent along main channels, where the manual centerline lies to the right in 83\% of cross sections, but not along tributaries (57\%), where it lies on the left in a majority of cross sections for T3 and T4 (Table S2). We consider possible causes in Section \ref{discussion_centerlines}.

\subsection{Edge detection: sensitivity and choice of settings}
\label{results_edges}

\begin{figure*}
\centering
\includegraphics[width=0.65\textwidth]{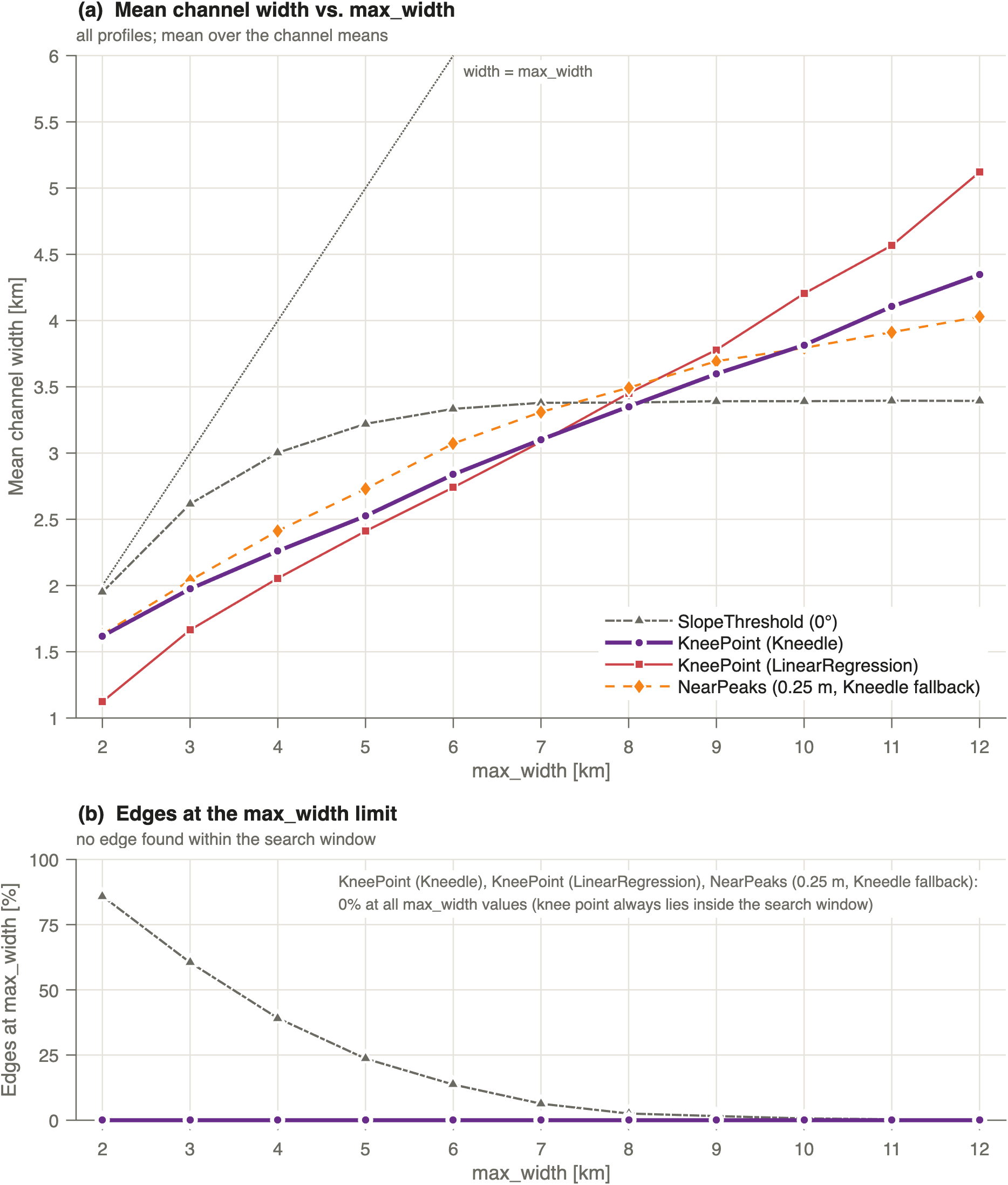}
\caption{Sensitivity of channel width to the maximum channel width of the edge search window (\texttt{max\_width}), for all 34 mapped channels on Roi Baudouin Ice Shelf (\texttt{min\_width} and other parameters fixed, as in Table \ref{tab:parameters}). (a) Mean channel width (mean of the per-channel mean widths); the dotted line marks width = \texttt{max\_width}. (b) Share of edges at the outer search limit, i.e., profiles on which no edge was found within the search window. Widths from the KneePoint methods increase steadily with \texttt{max\_width}, whereas SlopeThreshold widths level off once the search window is wide enough to include an appropriate edge; NearPeaks, which falls back to KneePoint-Kneedle where no prominent peak is found, lies in between. }
\label{max_width_sensitivity}
\end{figure*}

Channel widths depend strongly on the width of the edge search window (\texttt{max\_width}), and this dependence differs between methods (Fig. \ref{max_width_sensitivity}). SlopeThreshold widths increase only as long as the search window is too narrow to contain a point that meets the slope criterion: at a \texttt{max\_width} of 2 km, 86\% of its edges lie at the outer search limit, but fewer than 10\% from 7 km onward, beyond which its mean width levels off at $\sim$3.39 km (Fig. \ref{max_width_sensitivity}b). SlopeThreshold is instead sensitive to its threshold: at a \texttt{max\_width} of 8 km, changing \texttt{slope\_thr} by only $\pm$0.01$^\circ$ changes the mean width to 3.15 and 3.61 km, respectively (Fig. S5). For the KneePoint methods, in contrast, mean width increases steadily between 2 and 12 km and does not level off (Fig. \ref{max_width_sensitivity}a): from 1.62 to 4.35 km for Kneedle and from 1.12 to 5.12 km for LinearRegression. This is inherent to the knee point criterion, which evaluates the shape of the half-profile within the search window: widening the window changes the part of the profile that is considered (and, for Kneedle, the reference line from the centerline to the outer end of the window), and with it the location of the knee. As more of the surrounding ice shelf surface is included, the channel flank makes up a smaller part of the half-profile, and the knee tends to shift outward. NearPeaks behaves as a hybrid of the two (1.63 to 4.03 km). A topographic peak is a point where the outward slope drops to zero, so NearPeaks effectively applies the SlopeThreshold criterion at 0$^\circ$, with the additional requirement that the peak is sufficiently prominent. Where such a peak is found, the edge no longer changes once the window contains it; where none is found, NearPeaks falls back to Kneedle and inherits its dependence on \texttt{max\_width}. Its mean width therefore increases less steeply than that of Kneedle beyond $\sim$8 km (by 0.54 km between 8 and 12 km, compared with 1.00 km for Kneedle and 0.01 km for SlopeThreshold), but does not level off. 

Because the knee point has no fixed reference, \texttt{max\_width} has to be set deliberately for the KneePoint methods and, through its fallback, for NearPeaks. We use SlopeThreshold as a reference for this choice. Its criterion has a direct physical meaning at 0$^\circ$ (the point where the surface stops rising away from the centerline), and although its edges vary between individual profiles (see below), these errors act in both directions, so that its shelf-wide mean width is stable once the search window is wide enough. The mean widths of the KneePoint methods cross this stable value at a \texttt{max\_width} of $\sim$8 km (NearPeaks slightly earlier), which we therefore adopt as default. This choice calibrates the mean width of the KneePoint methods against that of SlopeThreshold. The close agreement of mean widths at 8 km (3.35 km for KneePoint-Kneedle, 3.45 km for KneePoint-LinearRegression, 3.49 km for NearPeaks and 3.38 km for SlopeThreshold, all within 0.14 km, compared with 0.29 km at 7 km and 0.39 km at 9 km) is therefore a consequence of this calibration rather than an independent confirmation of the edges. Of the KneePoint variants, we use Kneedle as default based on qualitative inspection of the returned edges (see next paragraph) and because its widths depend less strongly on \texttt{max\_width} than those of LinearRegression.

\begin{figure*}
\centering
\includegraphics[width=0.9\textwidth]{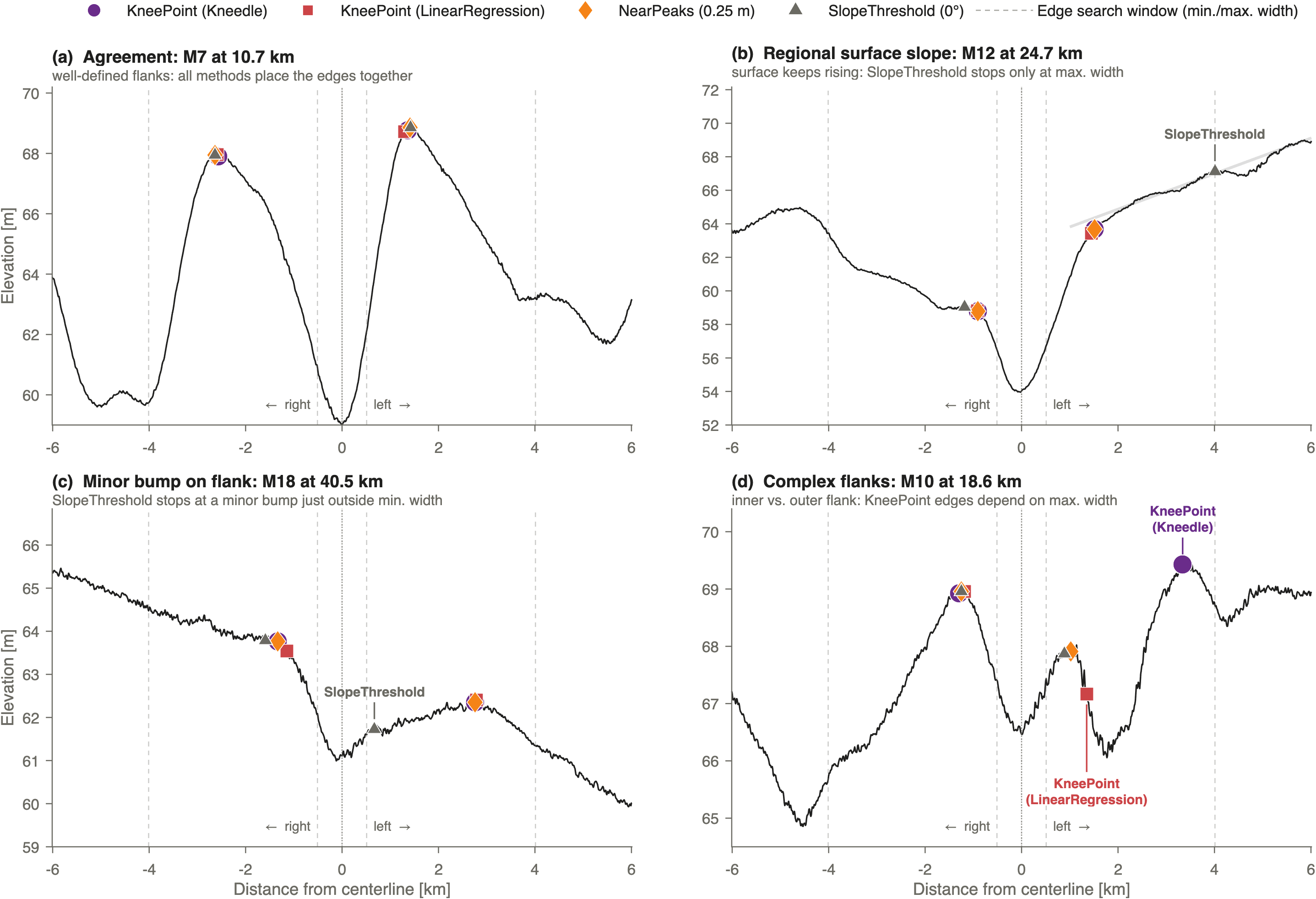}
\caption{Channel edges from the three edge detection methods (both KneePoint variants) on example cross sections (looking upstream; left edge at positive distances). (a) Well-defined flanks: all methods agree. (b) Regional surface slope (gray line): SlopeThreshold places the edge at \texttt{max\_width}, the KneePoint methods at the top of the channel flank. (c) A minor bump stops SlopeThreshold just outside \texttt{min\_width}. (d) Complex flanks with multiple ridges: Kneedle places the left edge on the outer ridge, the other methods on the inner one. Dashed lines: edge search window (\texttt{min\_width} = 1 km, \texttt{max\_width} = 8 km). The edges of all four method variants along every channel are compared in Figs. S42–S76.}
\label{edge_comparison}
\end{figure*}

The calibration constrains only the shelf-wide mean, and with \texttt{max\_width} fixed at 8 km, the methods can be compared on what it leaves unconstrained: the placement of edges on individual channels and profiles. Here, the methods agree for many profiles, but differ substantially for some. Per-channel mean widths from SlopeThreshold and KneePoint-Kneedle differ by a median of 0.39 km, and by more than 1 km for 11 of the 34 channels (Figs. S42–S76). SlopeThreshold is wider for about half of the channels and narrower for the rest, so that these differences largely cancel in the mean. They arise from the different criteria used by the methods. Where a channel runs on a regional surface slope, the outward slope may not drop below the threshold near the channel flank, and SlopeThreshold places the edge far out on the surrounding ice shelf or at \texttt{max\_width} (Fig. \ref{edge_comparison}b). Conversely, a minor ridge or bump on the half-profile can stop its search early, just outside \texttt{min\_width} (Fig. \ref{edge_comparison}c). The KneePoint methods instead evaluate the shape of the entire half-profile within the search window (Fig. \ref{edge_method_concepts}). Because both Kneedle and LinearRegression assess elevation relative to fitted or connecting lines rather than absolute slope, a uniform regional slope does not affect where they place the edge, and a minor bump carries little weight compared with the larger-scale channel flank. Where channel flanks are complex, with multiple ridges, the methods can select different edges (Fig. \ref{edge_comparison}d).

\subsection{Channel width, depth and asymmetry}

With the default configuration (KneePoint-Kneedle, \texttt{max\_width} = 8 km; Table \ref{tab:parameters}), the mean channel width is 3.35 ± 0.61 km (main channels 3.45 ± 0.67 km, tributaries 3.11 ± 0.38 km). Channel-averaged widths range from 2.34 km (M13) to 5.15 km (M24; Table S1). The mean channel depth, measured from the left edge to the centerline elevation, is 5.69 ± 2.93 m (main channels 5.43 ± 3.28 m, tributaries 6.32 ± 1.82 m), with channel-averaged depths ranging from 1.48 m (M23) to 12.78 m (M2). Width and depth also vary considerably along individual channels (Table S1; Figs. S7-S41), with typical within-channel standard deviations of 0.72 km and 3.03 m. Channels are moderately asymmetric: the left and right half widths (looking downstream) are similar (1.63 ± 0.36 km and 1.72 ± 0.45 km), but the left edge lies higher above the centerline than the right edge (5.69 ± 2.93 m vs. 4.53 ± 2.43 m). As a result, the left flank is on average steeper (3.90 ± 2.11 vs. 3.03 ± 1.88 m km$^{-1}$), and it is steeper in 65\% of cross sections. This asymmetry is consistent with Coriolis-enhanced melting along the left flank of Southern Hemisphere basal channels \citep{sergienko_basal_2013, gourmelen_channelized_2017, alley_evolution_2024}, and supports defining channel depth relative to the left edge.

\section{Discussion}

\subsection{Toward systematic and large-scale mapping}

Existing basal channel mapping, whether from optical imagery or DEMs, has relied largely on manual digitization, limiting the number of channels and the frequency at which they can practically be mapped. Sub-Ice substantially reduces the effort and cost of mapping channels by automating centerline mapping, profile extraction and edge detection. Because it applies the same criteria to every profile and channel, differences between channels mapped with a given configuration reflect differences in surface geometry rather than differences between investigators, whose influence has restricted comparisons between manually digitized records \citep{alley_evolution_2024}. 

Sub-Ice complements the automated approach of \cite{zinck_channelized_2026}, who extracted channel centerlines across nine ice shelves through watershed delineation of a downsampled DEM. Their approach requires no user-specified start and end points, which makes it well suited to detecting channels over large areas. Sub-Ice instead focuses on delineation: the user decides which channels to map, and the algorithm traces their centerlines at the resolution of the input DEM and delineates their edges, rather than representing channel width through a fitted curve. In that sense, Sub-Ice remains closer to manual mapping, while doing so efficiently and consistently. Its centerlines closely reproduce the manual record on Roi Baudouin Ice Shelf, whereas the agreement of watershed-derived centerlines with manual mapping \citep{alley_impacts_2016} varied between ice shelves, although the two comparisons rely on different reference datasets and ice shelves. Used together, the two approaches could combine large-scale channel detection with detailed delineation. 

Through mapping channels efficiently and consistently, Sub-Ice enables studies that have so far been impractical, including large-scale comparisons of channel width, depth, asymmetry and gradients within and among ice shelves in different oceanographic settings. Applied to time-stamped DEMs, Sub-Ice can also be used to analyze how these characteristics evolve over time and in response to changes in ocean or subglacial forcing, although this remains to be demonstrated. Whereas previous studies have necessarily focused on a small number of channels and studied them in detail, Sub-Ice is a step toward systematic, potentially Antarctic-wide mapping of basal channel geometry.

\subsection{Centerline accuracy and implications for manual records}
\label{discussion_centerlines}

Sub-Ice centerlines agree with the independent manual record of \cite{jackson_channelised_2026} with a median offset of 132 m, a small fraction of the mean channel width (3.35 km). This lies at the upper end of the differences among the manual records themselves: it equals the offset between the 2015 and 2021 records and exceeds that between records three years apart (74-96 m). A strict comparison with a single year is not possible, as the REMA mosaic combines acquisitions from several years. Moreover, the differences among manual records likely underestimate their uncertainty, because a displacement shared by all manual years (see below) cancels when they are compared. The remaining differences between Sub-Ice and manual centerlines are not random. First, where the two centerlines diverge, the Sub-Ice centerline almost always lies at a lower surface elevation (96\% of cross sections with offsets larger than 100 m). Second, offsets are largest along shallow channels. Third, the manual centerlines lie predominantly to the right of the Sub-Ice centerlines (74\% of cross sections, looking downstream, and 83\% along main channels), consistently across manual years. Together, these results suggest that much of the remaining disagreement reflects a displacement of the manual centerlines from the channel trough, rather than error in the Sub-Ice centerlines. 

We hypothesize that this displacement arises from how channels appear in optical imagery and how this appearance is interpreted during manual digitization. \cite{jackson_channelised_2026} digitized channels from the light-shadow contrast across their surface depressions in Landsat scenes, which arises from shading: the flank facing the sun appears brighter and the opposite flank darker than the surrounding surface. Due to Landsat's sun-synchronous orbit, Roi Baudouin Ice Shelf is consistently illuminated from the north-east. As the main channels run roughly from south to north, their right flank (looking downstream) faces west, away from the sun, and appears in every scene as a dark band extending up to several hundred meters from the trough. A centerline drawn along this band would lie on the right flank, consistent with the rightward displacement found in each manual year (70–75\% of cross sections) and along channels whose manual centerlines hardly moved between 2015 and 2021 (e.g., M8 and M18). Shallow channels, with their gentler flanks, show weaker contrast, which makes the trough harder to identify and its placement more interpretive. The digitizer found these shallow channels were the most difficult to map, consistent with the larger offsets found along them. Alternatively, an investigator placing the centerline between the visible edges of an asymmetric channel would also bias it toward the gentler, slightly wider right side. Other factors, such as a coregistration offset between the Landsat imagery and REMA, may also contribute. 

If confirmed, this bias has implications for existing basal channel records. Absolute channel positions digitized from optical imagery may be displaced from the channel trough by an amount on the order of 100 m, and by several hundred meters along shallow channels. Migration rates derived from scenes with similar illumination are likely less affected, as a consistent displacement largely cancels when positions are differenced. However, apparent migration could arise where a channel's depth, and thus its contrast, changes over time, or when records from sensors or scenes with different solar azimuths are combined. Channels with different orientations relative to the sun would also be displaced differently, which complicates comparisons between channels and ice shelves. The hypothesis could be tested by digitizing the same channels on scenes with contrasting solar azimuths, by comparing channels of different orientation, or by comparing manual centerlines with Sub-Ice centerlines derived from time-stamped DEM strips acquired close to the date of the imagery. Where the aim is to locate the channel trough, DEM-based mapping might be preferable.

\subsection{Definition of channel edges}
\label{discussion_edges}

Whereas Sub-Ice centerlines can be evaluated against an independent record, no such record exists for channel edges, and interpreting them is complicated by the lack of an agreed definition of a channel edge. Channel widths are regularly reported in the literature, implying that some criterion was applied, but this criterion is rarely made explicit and is often applied visually. A channel's edge is not a sharply defined feature: on most profiles the channel flank merges gradually into the surrounding ice shelf surface, cross-sectional shapes vary strongly between and along channels, and neighboring channels, fractures and other surface features complicate matters further. Where the edge is placed therefore depends on the criterion used, and widths and depths reported by different studies are difficult to compare.

Sub-Ice does not resolve this problem in an absolute sense: each of our three edge detection methods reflects its own assumptions about where exactly the channel edge lies rather than identifying a ``true'' physical boundary, and we have shown that results depend on the parameters set by the user (most notably \texttt{max\_width}, Fig. \ref{max_width_sensitivity}). What Sub-Ice does provide are explicit and reproducible criteria that are applied identically to every profile and channel. This allows channel edges to be compared meaningfully across the profiles and channels within a study and, with careful application, between studies. This also applies to dependent variables such as channel widths, edge elevations and therefore channel depths, as well as to their evolution along a channel and over time. 

Based on our sensitivity analysis and inspection of the detected edges (Section \ref{results_edges}), we recommend KneePoint-Kneedle as the default edge detection method. Compared with SlopeThreshold, it is less affected by regional surface slopes and minor profile features, and compared with KneePoint-LinearRegression, its widths depend less on the search window. Its main drawback is that it still depends on the search window: because the knee point has no fixed reference, \texttt{max\_width} effectively sets the scale of the mapped channel and must be chosen deliberately. This can be done as here, by choosing \texttt{max\_width} such that the mean Kneedle width matches the stable mean width of SlopeThreshold over the same channels. Alternatively, \texttt{max\_width} can be based on the typical spacing between neighboring channels, so that the search window spans the full channel cross section but not the flank of an adjacent channel. In either case, the edge method and \texttt{max\_width} should be kept fixed within a study and reported alongside the results. 

Whether a \texttt{max\_width} of 8 km is also suitable elsewhere remains to be tested. On Roi Baudouin Ice Shelf, channel-averaged widths of 2.34–5.15 km mean that an 8 km window spans the full cross section of each channel and part of the surrounding ice shelf surface. Comparable widths have been reported for channels on Cosgrove, Sulzberger and Ronne ice shelves \citep{alley_impacts_2016} and on Getz Ice Shelf \citep{chartrand_basal_2020}, and for modeled channels \citep{sergienko_basal_2013}, suggesting that a single search window of $\sim$8 km may be appropriate for many Antarctic basal channels. This would allow channel widths to be compared directly between ice shelves using a single configuration. Narrower channels, such as those near the grounding line of Ross Ice Shelf, which are about 1–2 km wide at the surface \citep{marsh_high_2016, whiteford_melting_2022}, may require a narrower window depending on the interplay with channel spacing. Whether the calibration against SlopeThreshold yields a similar window elsewhere, and how channel width and spacing affect the appropriate window, will need to be tested on an Antarctic-wide dataset.

Because they are difficult to define, channel edges have not been mapped in the existing literature in the way that channel centerlines have \citep[e.g.,][]{alley_impacts_2016}. Yet consistently defined edges open up questions that centerlines alone cannot address. Channel width and depth control where the ice is thinnest and how strain is concentrated, and thus the link between channels and basal or surface crevassing, weakened shear margins and transverse fractures that can precede calving \citep{vaughan_subglacial_2012, alley_troughs_2019, dow_basal_2018}. Systematic width and depth measurements across many channels, as facilitated by Sub-Ice, would allow these links to be examined statistically rather than case by case, and would provide observational targets for theories that predict channel width, asymmetry and spacing from plume dynamics and ice flow \citep{sergienko_basal_2013, drews_evolution_2015}.

\subsection{Surface expression vs. basal process}
\label{discussion_surface_vs_basal}

Sub-Ice maps the surface expression of basal channels, whereas channelized melt occurs at the ice-ocean interface, tens to hundreds of meters below. This raises the question of how far surface channel geometries can be used to infer basal geometries and processes. 

That basal channels are visible at the surface at all shows that their basal geometry is transmitted through the ice column, and where radar data are available, surface depressions identified as channels consistently correspond to channels at the ice shelf base \citep{alley_impacts_2016}. Under steady state, channels with a basal width greater than about five times the local ice thickness are expected to be fully hydrostatically adjusted \citep{drews_evolution_2015}. On Roi Baudouin Ice Shelf, where ice thickness near the channels is typically a few hundred meters \citep{paolo_widespread_2023}, the surface widths of nearly all channels mapped here comfortably exceed this threshold. As surface expressions are generally wider than the basal channels that produce them, this comparison is approximate, but it suggests that most of these channels are close to hydrostatic equilibrium. In that case, and assuming a constant firn air content across the channel, basal relief is about $\rho_i/(\rho_w-\rho_i) \approx 8$ times the surface relief, so that the mean surface depth of 5.69 m corresponds to a basal channel height of $\sim$45–50 m. Narrower channels can be held out of equilibrium by bridging stresses, in which case their basal relief is larger than their surface expression suggests. More generally, surface expressions are dampened reflections of basal geometries, and the ice shelf base is often more complex than the surface suggests \citep{langley_complex_2014}. Moreover, because channels can be advected for tens of kilometers without active melting, a surface channel does not by itself indicate active channelized melt at that location \citep{drews_evolution_2015}. Surface channel geometries are therefore most informative for wide, hydrostatically adjusted channels, and are best interpreted as indicators of basal geometry rather than direct measurements of it. 

However, the ability to map many channels efficiently and consistently partly offsets this limitation and creates an opportunity to learn more from this indirect surface signal: patterns that emerge across a large number of channels are far less likely to be single-channel artifacts. An example is the steeper left flank found here (3.90 vs. 3.03 m km$^{-1}$ on average, and steeper in 65\% of cross sections), which is consistent with Coriolis-enhanced melting along the left flank \citep{sergienko_basal_2013, gourmelen_channelized_2017, alley_evolution_2024}. Such patterns can support process-level inference in a way that case studies of individual channels or ice shelves cannot. The surface-based nature of our observations thus also presents an opportunity: whereas direct observations of conditions at the ice-ocean interface are sparse in space and time, high-resolution surface elevation data cover all Antarctic ice shelves over an increasingly long time span \citep{howat_rema_2022}. An important strength of Sub-Ice is that it can be applied to characterize basal channel geometry far beyond Roi Baudouin Ice Shelf and through time. Where available, radar and in-situ observations can then be used to relate surface and basal geometry.

\section{Conclusion and outlook}

We presented Sub-Ice, an open, semi-automated tool that maps the centerlines, cross-sectional profiles and edges of ice shelf basal channels from their surface expression in a DEM. Applied to Roi Baudouin Ice Shelf, it mapped 34 channels ($\sim$1,380 km), whose centerlines differ from an independent manual record by a median of 132 m, at the upper end of the differences between manual records from different years. Where the two diverge, Sub-Ice follows the channel trough more closely. Offsets are largest along shallow channels, and the manual centerlines are systematically displaced toward the right flank. We hypothesize that this displacement arises from the shading of channels in optical imagery and how it is interpreted during digitization. Edge detection methods based on different criteria can be calibrated to yield similar mean channel widths, but absolute widths, particularly for individual channels, depend on the method and the width of the edge search window. Keeping these settings fixed within a study and reporting them alongside the results keeps widths and depths comparable. Applied to time-stamped DEMs, Sub-Ice could track channel migration, widening and deepening over time, providing a consistent, DEM-based record of how channels respond to changes in ocean and subglacial forcing. 

Several of the limitations discussed above point to directions for future development rather than to inherent shortcomings of the approach. The largest remaining source of subjectivity is the manual selection of channel start and end points. This step could be automated by letting Sub-Ice determine the origin and terminus of a channel itself, for example from user-defined width or depth thresholds. Going further, Sub-Ice could scan a DEM for candidate channels directly by searching for elongated surface depressions that meet such criteria, possibly building on the watershed-based approach of \cite{zinck_channelized_2026}. This would reduce the risk of overlooking channels with a weak surface expression and make ``what counts as a channel'' explicit and reproducible. Similarly, the selection of a representative centerline from an ensemble could be automated using a quality criterion, and double channels could be detected and flagged from the cross-sectional profiles. Together, these developments would move Sub-Ice from a semi-automated toward a fully automated tool, capable of surveying entire ice shelves with minimal user interaction. 

Applying Sub-Ice beyond Roi Baudouin Ice Shelf would provide a consistent, Antarctic-wide record of basal channel geometry. It would also test whether, given differences in channel width and spacing between ice shelves, a single edge search window of $\sim$8 km is appropriate for many Antarctic basal channels, which would allow channel widths to be compared directly across ice shelves. Finally, although developed for basal channels, the approach of tracing topographic lows and delineating their edges could, with some adaptation, be applied to other elongated surface features relevant to ice shelf stability, such as rifts and large crevasses.

\section*{Supplementary material}
The Supplement contains the centerline ensembles (Figs. S1–S2), the sensitivity of channel widths to \texttt{sg\_window}, \texttt{min\_width}, \texttt{slope\_thr} and \texttt{peak\_prom} (Figs. S3–S5), per-channel geometry and centerline evaluation (Tables S1–S3), the manual centerline record (Fig. S6), a channel atlas (Figs. S7–S41) and a per-channel comparison of the edge detection methods (Figs. S42–S76).

\section*{Data availability statement}
Sub-Ice code and instructions are openly available at \url{https://github.com/DylanKreynen/Sub-Ice}. The version of Sub-Ice used for this study is archived at Zenodo (\citealp{kreynen_subice_2026}; \url{https://doi.org/10.5281/zenodo.24058679}). REMA surface elevation and MEaSUREs grounding zone data are publicly available from the Polar Geospatial Center Portal (\url{https://www.pgc.umn.edu}) and the National Snow and Ice Data Center (\url{https://nsidc.org}), respectively. Manually delineated channel centerlines are available in \cite{jackson_channelised_2026}. 

\section*{Acknowledgements} 
This work originated from a student project at the International Summer School in Glaciology in McCarthy, Alaska. 
KEA received funding from the Natural Sciences and Engineering Research Council (CAN) grant RGPIN-02910. 
MS received an Alaska summer school travel grant from the International Union of Geodesy and Geophysics (IUGG) and received funding from the SAGAZ project at Universidad de Magallanes, Chile.
RH received funding from the European Research Council (ERC) under the European Union’s Horizon Europe programme through the ERC-2022-ADG project GLACMASS (grant agreement No. 101096057). 
We thank A. Isaak for implementing the NearPeaks-Kneedle edge detection methods in Sub-Ice. 
Generative AI was used for debugging and refactoring Sub-Ice code as well as revising parts of the text in preparation for this submission. Tool used: Claude by Anthropic, models Sonnet and Opus, accessed between May and November 2026. 

\section*{Competing interests} 
The authors declare none.

\section*{Author contributions}
Sub-Ice was conceptualized by DK, MS and KEA; programmed by DK; tested by DK, MS and KEA; applied by DK. Manual centerlines were delineated by PHJ. The original draft was written by DK with input from RH and KEA; all authors contributed to editing and review. 

\bibliography{subicerefs} 
\bibliographystyle{igs} 

\end{document}